\documentclass[fleqn,10pt]{wlscirep}
\usepackage[utf8]{inputenc}
\usepackage[T1]{fontenc}
\usepackage{titlesec}
\usepackage{indentfirst}
\usepackage{amsmath}
\renewcommand{\vec}[1]{\mathbf{#1}}

\title{Deep unsupervised learning for Microscopy-Based Malaria detection}

\author{Alexander Tao, Boran Han \thanks{Harvard University, Cambridge, MA, US, email: boranhan.dl@gmail.com}}

\begin{document}
\flushbottom
\maketitle
\thispagestyle{empty}

\begin{flushleft}
{\bfseries\Large Abstract}
\\
Malaria, a mosquito-borne disease caused by a parasite,  kills over 1 million people globally each year. People, if left untreated, may develop severe complications, leading to death. Effective and accurate diagnosis is important for the management and control of malaria. Our research focuses on utilizing machine learning to improve the efficiency in Malaria diagnosis. We utilize a modified U-net architecture, as an unsupervised learning model, to conduct cell boundary detection. The blood cells infected by malaria are then identified in chromatic space by a Mahalanobis distance algorithm. Both the cell segmentation and Malaria detection process often requires intensive manual label, which we hope to eliminate via the unsupervised workflow.
\end{flushleft}

\section{Introduction}

Malaria, a mosquito-borne infectious disease, is widespread in tropical and subtropical regions around the equator. According to the 2019 World Malaria Report, malaria infected over 228 million people in 2018, with over 400,000 of those cases resulting in death. This disease brings with it an onslaught of symptoms, including fever, vomiting, and in extreme cases, coma. With a usual onset time of 10-15 days post exposure, various diagnostic methods are used to identify the malaria antigen, ranging from antibody detection to onsite clinical diagnoses, the traditional and most widely practiced method that consists of evaluating patients’ signs and symptoms and on physical findings at the exam\cite{OMS}. According to the CDC, malaria must be recognized promptly in order to treat patients in time and to also prevent the further spread of infection. However, they also note that the diagnosis of malaria can be difficult for various reasons: 1) Clinicians seeing malaria patients may forget to consider malaria among the potential diagnoses and laboratorians who examine blood smears under the microscope can fail to detect the parasites characteristic of the disease. 2) Technicians may be unfamiliar with, or lack experience with, malaria, and fail to detect parasites. In some areas affected by malaria, the disease is so potent that a large proportion of the population is infected by remain asymptomatic\cite{article1, WHO}. The current malaria diagnosis methods are not only labor costly, but they are also time consuming, which creates potential problems with regard to labor allocation of clinicians and economic considerations. In order to combat potential failures in diagnoses, we propose a deep learning and hue analysis method in order to render the diagnosing process free of human error. Therefore, automation on the diagnosis of malaria and other disease can potentially save the lives of the millions that are infected each year.

To improve the efficiency of malaria diagnosis, several attempts were performed using machine learning methods, trying to optimize different steps during diagnosis\cite{POOSTCHI201836, article10, article2, manescu2019deep, DBLP:journals/corr/abs-1804-09548, 8846750}. In the effort of improving the segmentation accuracy in thin blood smear, Rajaraman et al. compared different pre-trained CNN neural networks\cite{article2}. Manescu et al. developed a new deep learning method based on Convolutional Neural Networks (EDoF-CNN) to increase depth of focus in microscope, facilitating malaria detection\cite{manescu2019deep}. Later, Faster Region-based Convolutional Neural Network (Faster R-CNN), was proposed for cell segmentation, extraction of several single-cell features, followed by classification using random forests \cite{DBLP:journals/corr/abs-1804-09548}. However, all those methods mentioned above are supervised learning and it requires intensive manual label. Here we present a novel unsupervised learning method for malaria diagnosis, using deep learning with U-net architecture \cite{ronneberger2015unet} and hue analysis. 

With new hybrid loss function proposed in this paper, we present our method in malaria diagnosis using image set BBBC041v13\cite{article3}, available from the Broad Bioimage Benchmark Collection. In the method section, we introduce our deep learning and hue analysis method. In the results section, we present the results that we gathered from our learning. In the discussion section, we deliberate the implications and the logistics of our method.

\section{Methods}
\subsection{Malaria Detection Workflow}

\begin{figure}[h]
\centering
\includegraphics[width=0.85\columnwidth]{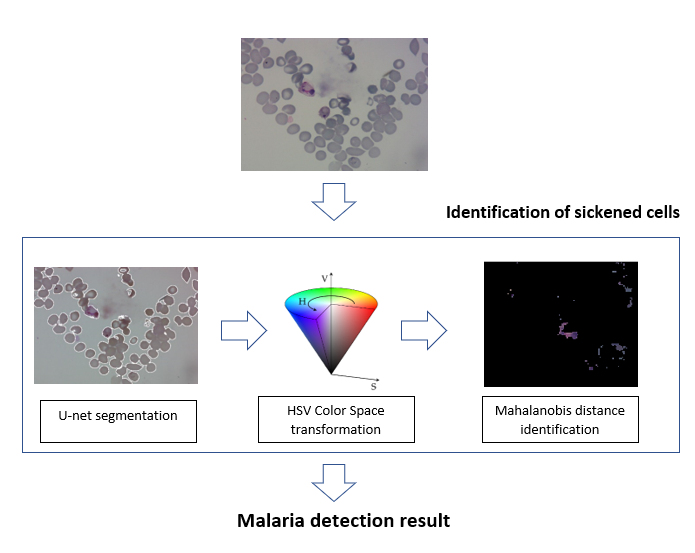}
\caption{Illustration of our proposed workflow.}
\end{figure} 

\bigskip
As shown in Fig. 1, we took one image as an example. First, we ran it through a preprocessing algorithm that normalizes the image so that the image can be fed into the U-net. The image is then passed through the U-net architecture where an unsupervised learning algorithm outputs a final image displaying cell boundaries. After this, the image is transformed into the HSV color
space to utilize hue analysis. At last, the Mahalanobis distance \cite{DEMAESSCHALCK20001} of each pixel, a multivariate measure that compares each pixel to the entire distribution, is then calculated and used to select out the malaria ridden cells.

\subsection{Self-training U-net architecture for cell boundary recognition}

Figure 2 depicts the U-net architecture used for the unsupervised cell boundary detection. The U-net architecture is separated into three distinct parts: The contracting/down-sampling path, bottleneck, and the expanding/upsampling path. The U-net structure is important to domains such as biomedical segmentation, since the number of samples, in this case, images of the patient’s blood cells, is usually limited. Both encoding and decoding paths of our U-net is composed of 4 blocks, with each block composed of 2 3x3 convolution layers and their respective activation functions (with batch normalization). There are two output of this U-net architecture: RMS output and SSIM output. The loss function used to minimize two outputs is described in the loss function section. Two outputs are added and convoluted with the Laplacian filter to be the final output. For the purpose of this study, the adaptive gradient decent algorithm called Adam\cite{article4} was used to train the model. The model was trained for 15 epochs with a learning rate of 0.01. 

\begin{figure}[ht]
\centering
\includegraphics[width=0.85\columnwidth]{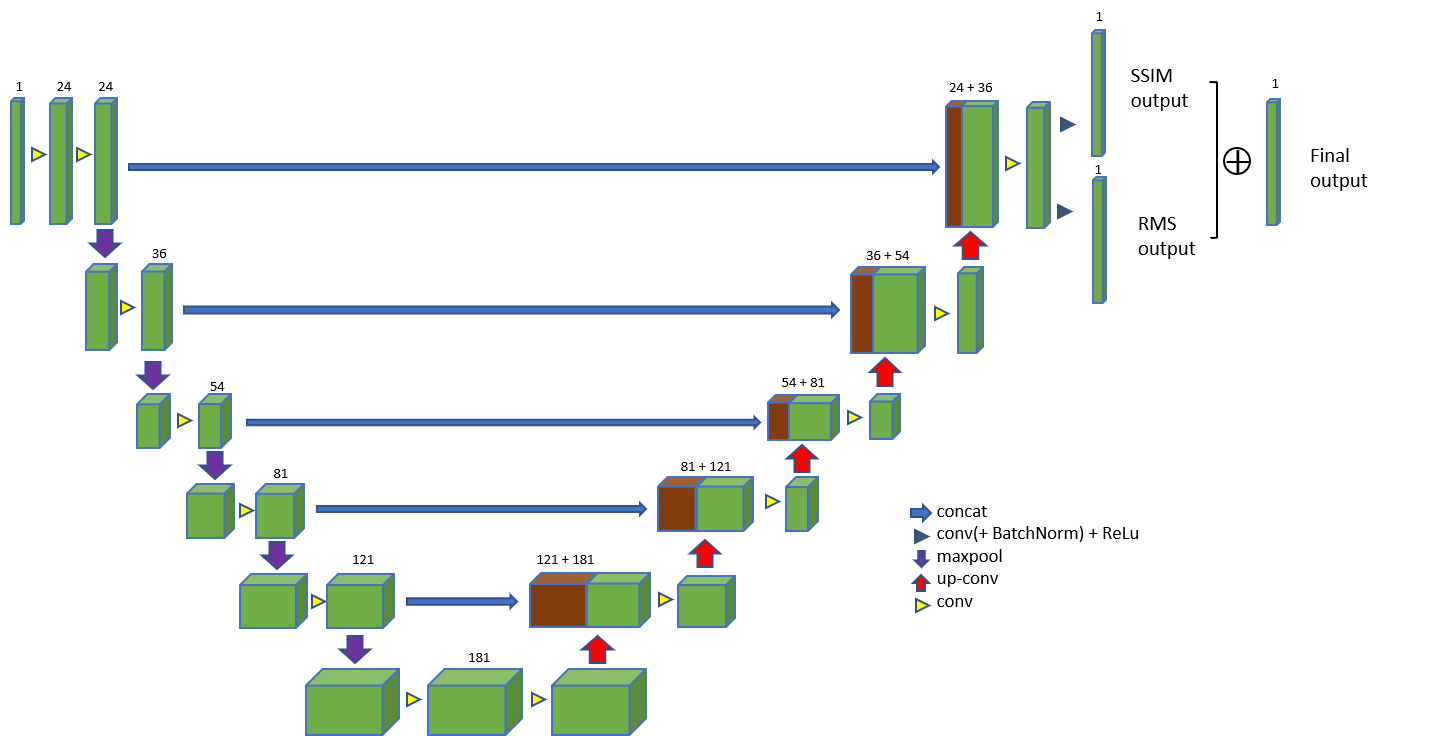}
\caption{U-net Architecture with two outputs: SSIM output and RMS output}
\end{figure} 

\medskip
The down-sampling path is represented by 5 layers of green blocks with convolution layers represented by the yellow arrows. Concatenation with corresponding cropped feature maps from the contracting paths are represented by the blue arrows and the max pooling process is depicted by the purple arrows. The bottleneck part of the network is between the contracting and expanding paths and is built from two convolutional layers with batch normalization and dropout. The up-sampling path is comprised of layers of green blocks with the deconvolution process represented by the red arrows. Concatenation with cropped feature maps from the contracting path is represented by the brown blocks. Two separate output segmentation maps are returned by the U-net.

\subsection{Hybrid loss function}

Our loss function $L$ consists of two parts: 1) Structural Similarity Index (SSIM) loss $L_{\text{SSIM}}$  that panelizes the structural difference\cite{1284395}, and 2) Root Mean Square (RMS) contrast loss $L_{\text{RMS}}$ that panelizes the contrast difference:

\begin{equation}
L= L_{\text{SSIM}}  +L_{\text{RMS}}
\end{equation}

$L_{\text{SSIM}}$, calculating the SSIM loss between SSIM output and input is defined as: 

\begin{equation}
L_{\text{SSIM}} = 1 -  \frac{1}{N}\sum_{i = 0}^N f_{\text{SSIM}}( \nabla\hat{y}^{(i)}_\text{SSIM},\nabla{X^{(i)}})
\end{equation}

Where $X$ is the input of the U-net output, N is the batch size, \textit{i} represents the \textit{i}th in each batch, $\nabla$ is the gradient operator and $f_{SSIM}$ is the function calculating SSIM between two tensors, described by Wang, et. al\cite{1284395}. This SSIM loss will ensure the output to have similar structure with the input microscopy images.

Meanwhile, to segment the image, we further added a $L_\text{RMS}$. It is set as the mean squared loss of the RMS contrast between $\hat{y}_\text{RME}$ and binarized $X$

\begin{equation}
L_\text{RMS} = \frac{1}{N}\sum_{i = 0}^N(h_\text{RMS}(\hat{y}^{(i)}_\text{RMS}) - h_\text{RMS}(O( X^{(i)}))^{2}
\end{equation}

Where $O$ is the function binarizing the $i$th image of $X^{(i)}$ using Otsu’s method \cite{4310076}, and $h_\text{RMS}$ is a function measuring the image contrast, equivalent as the standard deviation of the pixel intensities\cite{Peli:90}:

\begin{equation}
h_\text{RMS}(I)= \sqrt{ \frac{1}{C_{1}C_{2}} \sum_{i=0}^{C_1} \sum_{j=0}^{C_2}(I_{ij}-\bar{I}})^{2}
\end{equation}

where intensities $I_{ij}$ are the pixel value at position $(i,j)$ of an image. $C_{1}$,$C_{2}$ are the image size in two dimensions. $\bar{I}$ is the average intensity of all pixel values in the image. The RMS loss ensures that the U-net output images has the contrast has its corresponding binarized images. Please note that although binarized image was used in the loss function, it is only used for U-net to “learn” the contrast difference between the $\hat{y}^{(\textit{i})}_\text{RMS}$ and $\textit{O}\textit(X)^{(\textit{i})}$, instead of pixel-wise intensity difference.

\subsection{Malaria detection in chromaticity space}

Second, we used the hue analysis to select the infected red blood cell. We transformed the RGB color space to an HSV color model. This model, a conical representation of color in 3D space separates chromaticity and easily allowed us to detect and identify the malaria infected cells. Based off the results of our hue detection algorithm, we were able to determine whether the given sample was malaria infected.
\medskip
The following equations were utilized to transform RGB into the HSV color space:

\begin{align}
\label{eq:hsv}
 H &=\begin{cases}0 & B\leq G\\360- \theta  &  B\geq  G\end{cases} \nonumber\\
 \theta & = \cos^{-1}\{{ \frac{ \frac{1}{2}\times[(R -G) + (R-B)] }{[(R-G)^{2}+ (R-B)(G-B)] } }\} \\
 S &= 1 -  \frac{3}{(R+G+B)}[\min(R,G,B)] \nonumber \\
 V &= \frac{1}{3}(R + G + B) \nonumber 
\end{align}

In order to separate the outlier color pixels from the rest of the image, we utilized Mahalanobis distance \cite{DEMAESSCHALCK20001}, a multivariate distance metric that allowed us to measure the distance between a vector and a distribution. Euclidean distance, a more commonly used metric, was not used because of its inability to consider how the rest of the points in a dataset vary beside the two points being compared. Instead, Mahalanobis distance allows us to calculate an accurate representation of a point’s relation to the entire distribution. It is calculated in three steps. First, the columns are transformed into uncorrelated variables, then, they are scaled so that their variance is equal to one. Finally, it calculates the Euclidean distance between the points after scaling. The formula to compute this is as follows:
\begin{equation}
D^{2}(\vec x) =(\vec x-\vec m)^{T} \vec C^{-1}(\vec x-\vec m)
\end{equation}
Where $D^2$ is the square of the Mahalanobis distance, $\vec x$ is the dataset with two features, which is the $H$ and $S$ of each pixel defined in (\ref{eq:hsv}), $m$ is the mean values of each feature , and $C^{-1}$  is the inverse of the covariance matrix of the independent variables. Assuming that the test statistic follows chi-square distributed with 2 degree of freedom ($H$ and $S$), the critical value at a 0.01 significance level set as a threshold, which means that pixel can be considered as extreme (Malaria) if $p<0.01$.

\begin{figure}[h]
\centering
\includegraphics[width=0.65\columnwidth]{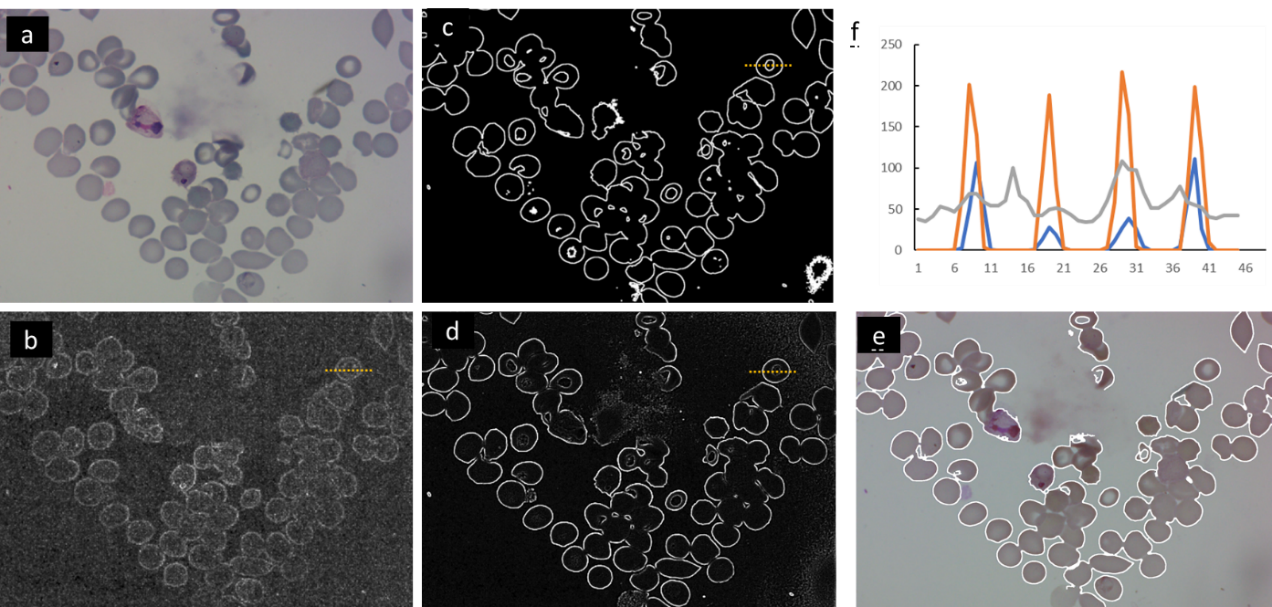}
\caption{Results of cell boundary detection. (a) raw image. (b) boundary detection using Laplacian operator, (c) Sobel operator
on binarized image and (d) prediction from self-supervised U-net model. (e) raw image overlaid with prediction from U-net. (f)
intensity projection of the cross section indicated as the yellow dashed lines in (a) – (d).}
\end{figure}

\begin{figure*}[h]
\centering
\includegraphics[width=0.65\columnwidth]{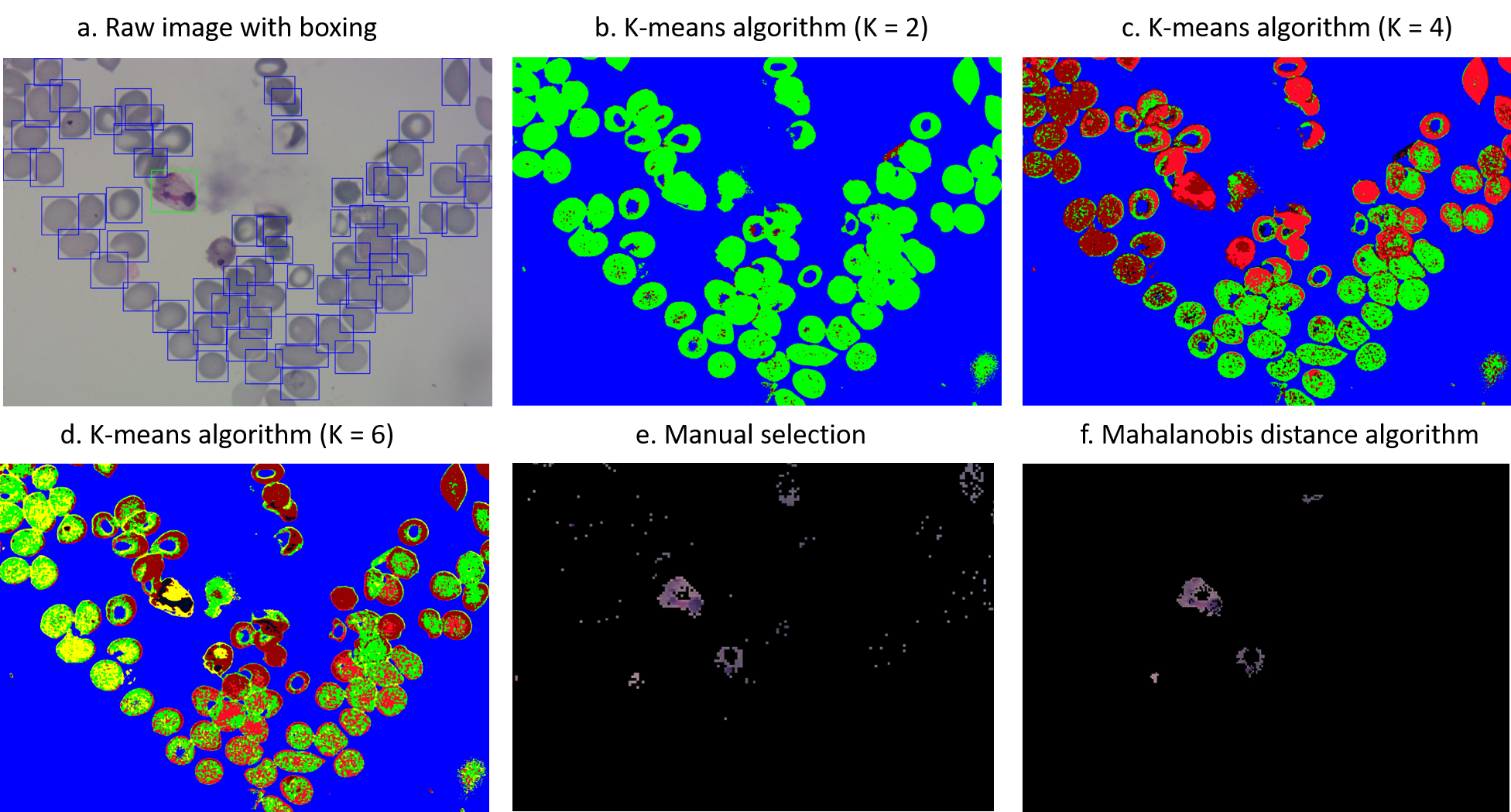}
\caption{Results of Hue Analysis using mahalanobis distance}
\end{figure*}

\section{Results}

First, we used our trained U-net model to predict the cell boundary, which is described as the first step in Figure 1. The prediction results are shown in Figure 3d and 3e. To compare with other popular methods, we also performed a Laplacian operator directly on the raw image (Figure 3b) and on binarized image (Figure 3c). Despite both methods being able to provide cell boundary, the quality of the output is degraded. To further quantify our findings. We plotted intensity of the marked cross section for all the methods above (Figure 3f). We found that using Laplacian operator directly on raw image, the signal can easily be mask by the large noise; using Laplacian operator on the binarized image, unsharpened edge inside cell can be selected too. In contrast, our U-net model with hybrid loss function finds a balance between these two methods: it shows strong signals detecting cell boundary, and weaker signals inside cell. 

\begin{figure}[h!]
\centering
\includegraphics[width=0.65\columnwidth]{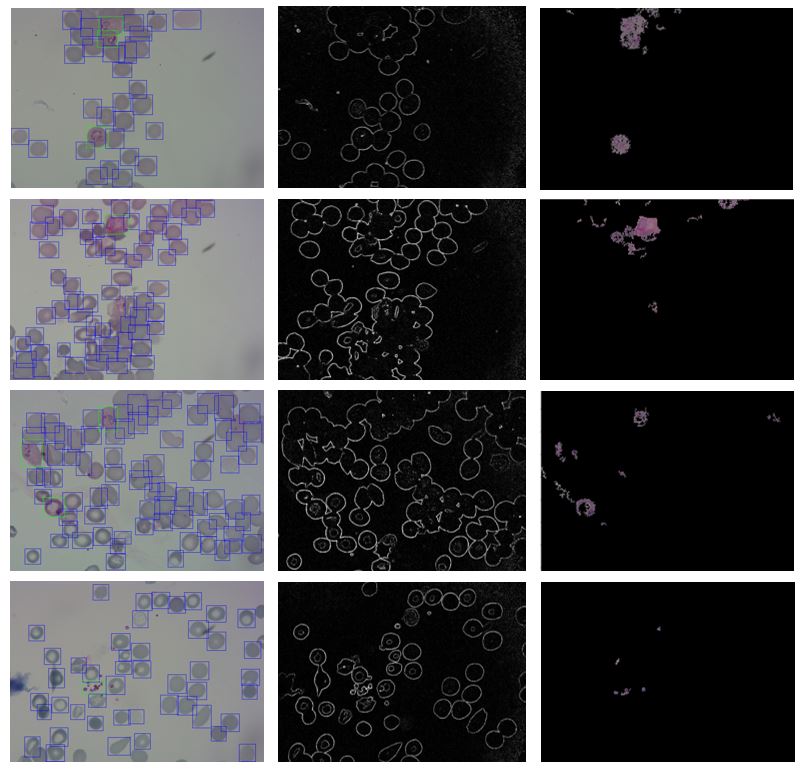}
\caption{The images above are the results of running our U-net and Mahalanobis algorithm on other images. The left column contains raw images, with blue boxing around normal red blood cells and green boxing around malaria infected ones. The center column contains the results of unsupervised U-net algorithm. The final column contains the results of running the images through our Mahalanobis distance and connected components algorithm.}
\end{figure}

Figure 4 depicts unique hue analysis approaches that we used in our malaria detection process. The first image, raw image with boxing, shows the malaria ridden cell boxed in green. The next three images are the results of the utilization of a K-means algorithm with different values of K. However, malaria ridden cells were not selected out, which means K-mean is not an efficient algorithm for automatic malaria detection. Manual selection, the second technique that we used, involved setting specific ranges of values for the hue and manually changing them until the most accurate identification of malaria was reached. After fine tuning, we set a high boundary for HSV selection of (180, 180, 255) and a low boundary of (128, 40, 0). Our last and most efficient technique, utilizing a Mahalanobis distance algorithm along with a connected components algorithm, is depicted in the last box. The connected components algorithm, utilized to get rid of both noise and cells unrelated to the malaria-positive red blood cell, labels subsets of connected components based on a manually set pixel blob size. The algorithm traverses the pixels in the image, labeling pixel vertices based on their connectivity and relative values of neighboring pixels. Following this labeling stage, the graph is divided into subsets. Regions of interest in these divided images are then able to become isolated. Compared with our last techniques, manual selection of color range is more accurate, however our proposed method is a parameter-free method, which means it requires no manual parameter tuning. Figure 5 shows more examples of malaria detection using our proposed workflow. 

\section*{Conclusion}

Malaria is no doubt among one of the most harmful disease in the developing world. Nearly half of the world’s population spread across 91 different countries are at risk of malaria transmission. Therefore, an efficient and accurate diagnoses of it is vital to improving the survivability rate of the illness. In recent years, methods utilizing artificial intelligence have had aspects of manual selection and manual identification. Through this project however, we demonstrate a more efficient automated malaria identification process that utilizes hue analysis to pinpoint malaria ridden cells. At this moment in time, the memory allocation of CPU and GPU for Android and IOS machine learning libraries is sufficient, meaning that the computation cost of our research should not be an impediment. The model that we have proposed only takes up (insert amount) MB and took little RAM to process test data. The model could serve as a tool to help single out malaria ridden red blood cells when fed pictures of a patient’s blood. It could minimize delays and improve accuracy in malaria diagnosis.

\bibliography{main}

\end{document}